\documentclass[twocolumn]{aastex62}

\shorttitle{Blueprint for the Milky Way's Stellar Populations}
\shortauthors{An \& Beers}

\accepted{for publication in the Astrophysical Journal after minor revision}

\begin{document}

\title{A Blueprint for the Milky Way's Stellar Populations: The Power of Large Photometric and Astrometric Surveys}

\author{Deokkeun An}
\affiliation{Department of Science Education, Ewha Womans University, 52 Ewhayeodae-gil, Seodaemun-gu, Seoul 03760, Republic of Korea; deokkeun@ewha.ac.kr}

\author{Timothy C.\ Beers}
\affiliation{Department of Physics and JINA Center for the Evolution of the Elements, University of Notre Dame, Notre Dame, IN 46556, USA}

\begin{abstract}

Recent advances from astronomical surveys have revealed spatial, chemical, and kinematical inhomogeneities in the inner region of the stellar halo of the Milky Way Galaxy. In particular, large spectroscopic surveys, combined with Gaia astrometric data, have provided powerful tools for analyzing the detailed abundances and accurate kinematics for individual stars. Despite these noteworthy efforts, however, spectroscopic samples are typically limited by the numbers of stars considered; their analysis and interpretation are also hampered by the complex selection functions that are often employed. Here we present a powerful alternative approach -- a synoptic view of the spatial, chemical, and kinematical distributions of stars in the Milky Way based on large photometric survey databases, enabled by a well-calibrated technique for obtaining individual stellar metal abundances from broad-band photometry. We combine metallicities with accurate proper motions from the Gaia mission along the Prime Meridian of the Galaxy, and find that various stellar components are clearly separated from each other in the metallicity versus rotation-velocity space. The observed metallicity distribution of the inner-halo stars deviates from the traditional single-peaked distribution, and exhibits complex substructures comprising varying contributions from individual stellar populations, sometimes with striking double peaks at low metallicities. The substructures revealed from our less-biased, comprehensive maps demonstrate the clear advantages of this approach, which can be built upon by future mixed-band and broad-band photometric surveys, and used as a blueprint for identifying the stars of greatest interest for upcoming spectroscopic studies.

\end{abstract}


\section{Introduction}

The inner region of the Milky Way's stellar halo has long been thought to be composed of a single stellar population, characterized by old ages and low metallicities with a single abundance peak at ${\rm [Fe/H]} \approx -1.6$ \citep{ryan:91}, along with a small net rotational velocity around the Galactic center. However, this simple picture has been radically altered in recent years, thanks to the combined efforts of large astronomical surveys in both the optical and near-infrared spectral regions. Targeted spectroscopic observations of individual stars provide a wealth of information on their elemental abundances, and radial velocities derived from such spectra can be combined with proper motions to compute a full space-velocity vector.

In this way, previous analyses of medium-resolution spectra from the Sloan Digital Sky Survey \citep[SDSS;][]{abolfathi:18} led to the suggestion that the local stellar halo is dominated by two spatially overlapping populations of stars, known as the inner-halo (IH) and outer-halo (OH) components, respectively \citep{carollo:07,beers:12}. In addition, accurate parallaxes and three-dimensional motions derived from the Gaia mission indicated that a large fraction of the local halo stars belong to a proposed progenitor dwarf galaxy (called Gaia-Enceladus; GE or Gaia-Sausage) from a single massive past accretion event in the formation history of the Milky Way \citep{belokurov:18,helmi:18}. Furthermore, a large number of metal-rich ([Fe/H] $>-1$) stars with halo kinematics (referred to as the Splashed Disk; SD) \citep{bonaca:17,belokurov:19,dimatteo:19,amarante:20}, and a group of metal-poor stars with disk-like kinematics (referred to as the Metal-Weak Thick Disk; MWTD) \citep{morrison:90,beers:14,carollo:19} have been discovered, adding complexity to the traditional three-component (thin disk, thick disk; TD, and halo) model of Galactic stellar populations \citep[e.g.,][]{ivezic:08}.

However, confirmation of each component as distinct entities, and exploration of their inter-relationships, are challenging because these studies are based to a large extent on spectroscopic data, which inherit often-complex target-selection functions. Consequently, piecewise information on individual components (often from multiple surveys with differing selection functions) must be stitched together in order to reconstruct the multi-dimensional structure of the Milky Way's stellar populations.

In contrast, multi-filter photometric imaging data are far less affected by sampling biases, and, when properly calibrated to stars of known metallicities, can be used to obtain an unbiased view of the nature of stellar populations over an enormous volume of the Galaxy, providing significantly larger samples for detailed analysis and future spectroscopic exploration. In this paper, we present a multi-dimensional map of stars in the inner stellar halo based on large photometric survey data. Sample selections are described in \S~\ref{sec:sample}. Distance and metallicity estimations are briefly summarized in \S~\ref{sec:method}. \S~\ref{sec:results} presents a blueprint of the Milky Way's halo, which clearly identifies and distinguishes previously suggested stellar populations in the metallicity versus Galactocentric rotational velocity space. Our findings are discussed in the context of the Galaxy's formation in \S~\ref{sec:discussion}.

\section{Photometric and astrometric data}\label{sec:sample}

The SDSS Legacy imaging survey currently provides the most uniform, wide, and deep photometric data over a large range of wavelengths in five broad-band filters ($ugriz$). We made use of broad-band photometry in the $ugriz$ passbands from the 14th Data Release (DR14) of the SDSS IV \citep{abolfathi:18}. This version of photometry is based on the photometric calibration called a hyper-calibration procedure \citep{finkbeiner:16}, which uses Pan-STARRS1 photometry in $griz$ \citep{schlafly:12} to minimize global zero-point offsets throughout the survey area.

The original SDSS data acquisition was performed in a drift-scan or time-delay-and-integrate (TDI) mode, which took multi-band images simultaneously with the same effective exposure time of $54.1$~sec. As a result, the signal-to-noise ratio is lower in the $u$ passband due to lower quantum efficiency in the detector at this wavelength, which is also where most stars emit less flux than in the $griz$ passbands. Since the SDSS $u$-band data are not sufficiently deep to make full use of the photometry in $griz$, we have constructed a secondary data set, supplementing SDSS $griz$ data with deep $u$-band photometry from the South Galactic Cap of the $u$-band Sky Survey \citep[SCUSS;][]{gu:15}. The SCUSS photometry is at least one magnitude deeper than the SDSS $u$ images, and therefore can be used to obtain more precise metallicity estimates. The filter response function of the SCUSS $u$ has a narrower edge on the long wavelength side ($\sim3800$~\AA) than the SDSS $u$, which introduces small corrections in the magnitude conversion \citep{gu:15}. Using accurate photometry ($\sigma < 0.02$~mag) of point sources at high Galactic latitude ($|b| > 60\arcdeg$) that are provided in both catalogs, we derived a transformation equation between the two systems:
\begin{eqnarray}
u_{SDSS} = u_{SCUSS} + 0.0085 - 0.0435 (u_{SCUSS} - g_{SDSS}) \nonumber \\
+ 0.0549 (u_{SCUSS} - g_{SDSS})^2 - 0.0163 (u_{SCUSS} - g_{SDSS})^3, \nonumber \\
\end{eqnarray}
where the subscript indicates each of the photometric systems.
 
\begin{figure*}
\centering
\includegraphics[scale=0.86]{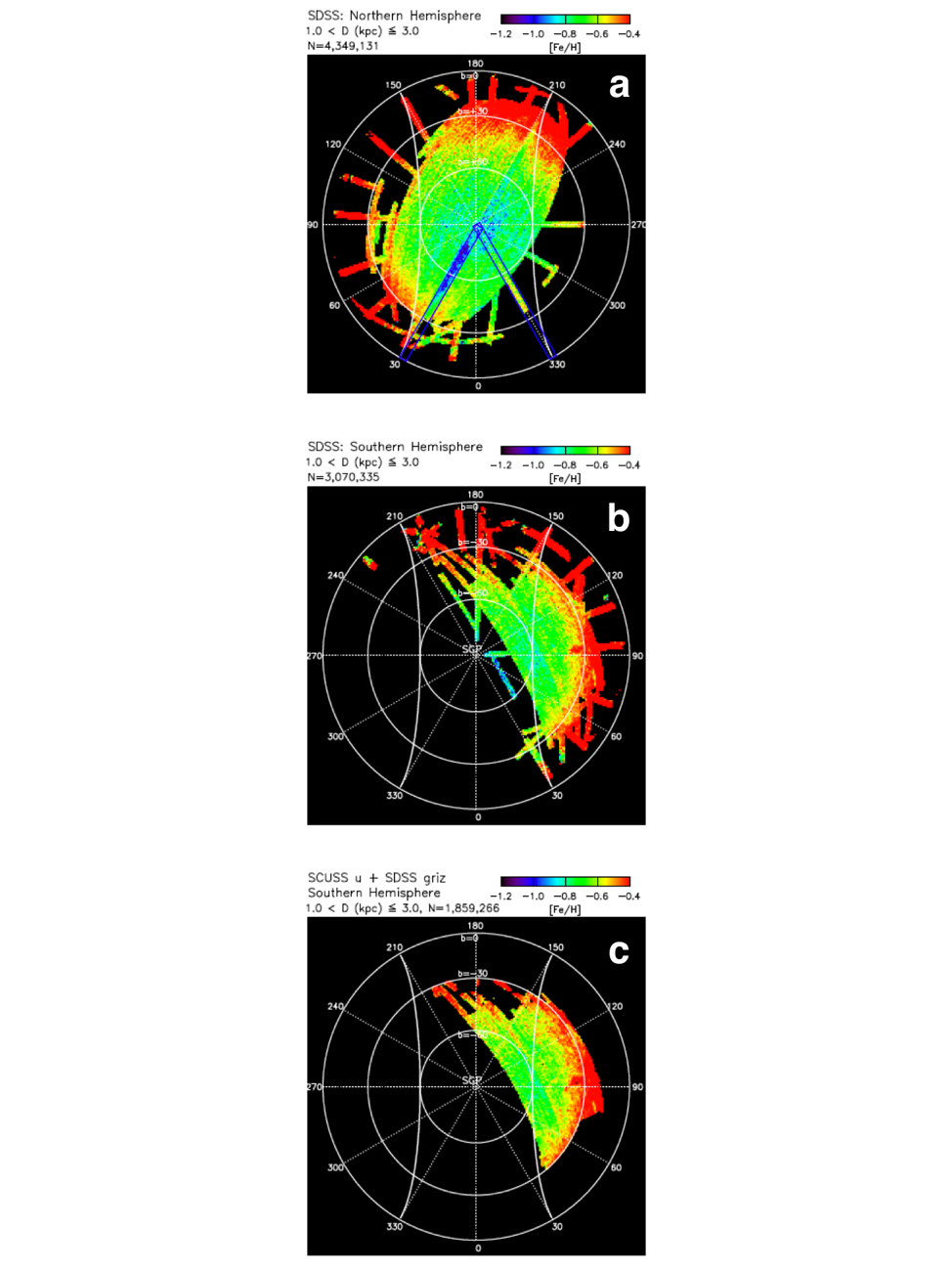}
\caption{Survey regions and sample selection. Sky coverage of the SDSS imaging survey in the Northern (panel a) and Southern Galactic Hemispheres (panel b). The median metallicity of stars at distances $1 < D ({\rm kpc}) < 3$ from the Sun along each line of sight is shown by different colors, computed from a generalized (error-weighted) metallicity distribution. Stars in the rectangular regions at $l = 30\arcdeg$ and $l = 330\arcdeg$ in panel (a) are not included in this work, due to potentially large systematic errors in their photometry.  Panel (c) shows the sky coverage from the SCUSS imaging in the Southern Galactic Hemisphere. Metallicities are estimated from the SCUSS $u$-band photometry and SDSS $griz$ data. In all panels, the concave-lens-shaped region depicts a field orthogonal to Galactic rotation, within which we select our sample, defined as a region of width $\pm30\arcdeg$ along the Galactic Prime Meridian, a great circle at $l = 0\arcdeg$ and $l = 180\arcdeg$.}
\label{fig:map} \end{figure*}

Figure~\ref{fig:map} shows survey regions in the Galactic coordinate system. The SDSS images cover both the Northern and Southern Galactic Hemispheres, while the combined SCUSS/SDSS catalog only covers the Southern Galactic Hemisphere. We excluded photometric measurements along the narrow strips centered at $l = 30\arcdeg$ and $l = 330\arcdeg$, where photometrically derived metallicities are systematically higher than in surrounding areas, possibly due to a small calibration error in the SDSS imaging strips. In all cases, we restricted our sample to the Galactic Prime Meridian (a $\pm30\arcdeg$ strip centered at Galactic longitudes $l =0\arcdeg$ and $l = 180\arcdeg$), where rotational velocities ($v_\phi$) of individual stars in the Galactocentric cylindrical coordinate system can be obtained with proper-motion data alone (see below), without requiring spectroscopic radial-velocity measurements \citep[e.g.,][]{bond:10}.

We chose a $1$ arcsec search radius to cross-match photometric sources with those in the DR2 catalog of the Gaia mission \citep{gaia:18}. We imposed an upper limit on the proper-motion errors $\sigma_\pi/\pi < 0.3$ on each direction in celestial coordinates. We also corrected parallaxes for the global parallax zero-point offset (0.029 mas) as suggested by the Gaia team \citep{lindegren:18}.

\section{Method}\label{sec:method}

\subsection{Determination of Stellar Parameters}

For each set of photometric catalogs (SDSS $ugriz$ or SDSS $griz$+SCUSS $u$), we estimated distance, metallicity, and mass (or effective temperature, $T_{\rm eff}$) for each star by searching for the best-fitting model to the $ugriz$ data \citep{an:13,an:15a,an:19}. These models were originally built from the Yale Rotating Evolutionary Code \citep{sills:00} and semi-empirical color-$T_{\rm eff}$ relations \citep{lejeune:97}, but showed deviations from the observed colors of main-sequence stars in a number of well-studied star clusters \citep{an:08}. The differences are of the order of a few hundredths of magnitude, which vary systematically as a function of $T_{\rm eff}$. To remove the systematic trends with $T_{\rm eff}$, \citet{an:09,an:13} obtained color-$T_{\rm eff}$ corrections to the models to match the observed main sequences of well-studied star clusters over the metallicity range $-2.4 \leq {\rm [Fe/H]} \leq +0.4$; see \citet{an:07,an:15b} for more information on the basis of the empirical color-$T_{\rm eff}$ correction procedure. As shown below, photometric metallicities derived using the empirically calibrated isochrones are as precise as 0.3 dex for bright stars, with additional systematic errors of the same order, but relative metallicity estimates are more robustly predicted.

We adopted foreground extinction along each line of sight from a dust emission map \citep{schlegel:98}. The SCUSS $u$ data were transformed before applying the extinction corrections. We limited our analysis to stars with small foreground reddening, $E(B\, -\, V) < 0.1$, at high Galactic latitude, $|b| > 20\arcdeg$, to avoid potentially large systematic errors in the extinction measurements and/or in the extinction coefficients \citep{schlafly:11}.

We also restricted our sample to those stars having $4.5 < M_r < 7.5$~mag and $u < 20$~mag (or $u < 21$ in SCUSS), along with $\sigma_{\rm [Fe/H]} < 1.5$~dex (note that the typical error in photometric metallicity estimates for our stars is considerably lower, on the order of 0.3 dex for bright stars, and that the effect of stars with large photometric metallicity errors is mitigated by our adopted weighting scheme, which goes as $1/\sigma_{\rm [Fe/H]}^2$) and $\chi^2 < 5$ from all five passbands, where $\chi^2$ is a total chi-square value from the best-fitting model \citep[see][for more information]{an:13}.

Although photometric samples generally suffer significantly less bias than spectroscopic samples, they are not entirely free from it either, since any apparent magnitude and/or color cuts can disfavor stars of certain types or physical properties. For example, in the context of bias against metallicity, metal-rich main-sequence stars are under-populated at large distances, because they are intrinsically fainter than their metal-poor counterparts. Lower main-sequence stars are also more difficult to include in distant halo samples, due to their low luminosities. All of these effects can lead to a metallicity distribution skewed toward higher metallicities. In our previous work \citep{an:13}, we avoided this sampling bias by applying a strong selection condition based on inferred stellar mass, in order to construct a volume-limited sample that included distant halo stars. In this work, however, we adopted a minimal sample selection criterion based on M$_r$ (to avoid bias toward more massive, metal-rich main-sequence stars) and restricted our analysis to stars with well-measured $u$-band photometry, which is essential to derive reliable photometric metallicities. Consequently, we do not seek to assign quantitative estimates of the fractions of the various populations that are identified, but rather simply point out their existence. Full assessment of the effects of any remaining sample bias will be presented in a subsequent paper.

\subsection{Verification of Stellar-Parameter Estimates}

\begin{figure*}
\centering
\includegraphics[scale=0.96]{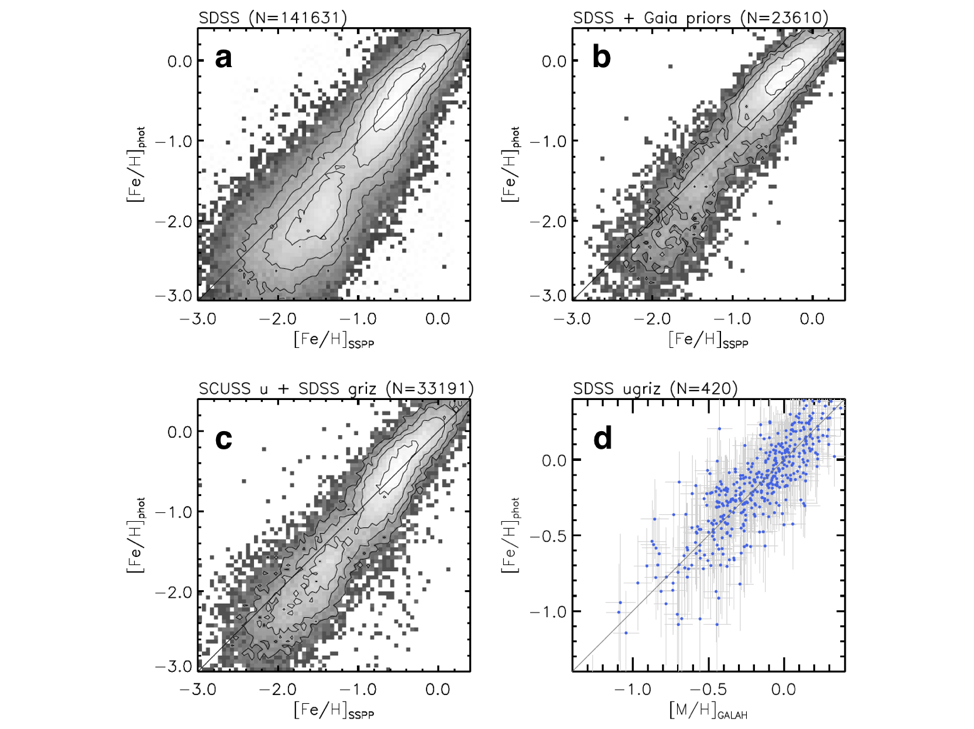}
\caption{Comparison between photometric and spectroscopic metallicity estimates. Panel (a): Comparison between photometric and spectroscopic metallicities from the medium-resolution spectra in SDSS for main-sequence dwarfs. Photometric metallicities are based on the SDSS $ugriz$ data. The gray scale represents a logarithmic number density of the stars; the contours indicate $20$, $60$, $85$, and $95$ percentiles. Panel (b): Same as in panel (a), but based on photometric metallicity estimates with Gaia priors. Panel (c): Same as in panel (a), but based on SCUSS $u$-band photometry in the metallicity estimation. Panel (d): Comparison between photometric and spectroscopic metallicities from high-resolution spectra in the GALAH survey. Only dwarfs are included in the GALAH comparison, for which the selection is limited for the metallicity comparison to [Fe/H]$\ge-1$.}
\label{fig:param} \end{figure*}

Panel (a) of Figure~\ref{fig:param} shows a comparison of photometric metallicities with spectroscopic estimates from the medium-resolution spectra in SDSS, which have been analyzed using the SEGUE Stellar Parameter Pipeline \citep[SSPP;][]{lee:08}. Overall, the agreement is satisfactory, although there are systematic departures seen in the metal-rich and the metal-poor sides from the full photometric solutions. The offset amounts to $\sim 0.3$~dex at [Fe/H]$ = -2$, in the sense that photometric metallicities are lower, and to $\sim 0.2$~dex for metal-rich stars ([Fe/H]$ > -1$). The same systematic departures are seen in the case of photometric metallicities with Gaia priors on distance (panel b), and in the case of using SCUSS $u$-band photometry (panel c). However, we did not correct photometric metallicities to match the SSPP scale, since the latter is also subject to calibration using high-resolution spectroscopic analysis \citep{hayes:18}. Reassuringly, the comparison with high-resolution data for metal-rich dwarfs from the GALAH survey \citep{desilva:15,buder:18} exhibits no noticeable systematic differences for metal-rich stars (panel d). Since the SSPP sample does not uniformly cover the entire stellar-parameter space, the systematic differences may reflect problems in stellar models at certain surface temperatures or gravities of stars.

\begin{figure}
\centering
\includegraphics[scale=0.9]{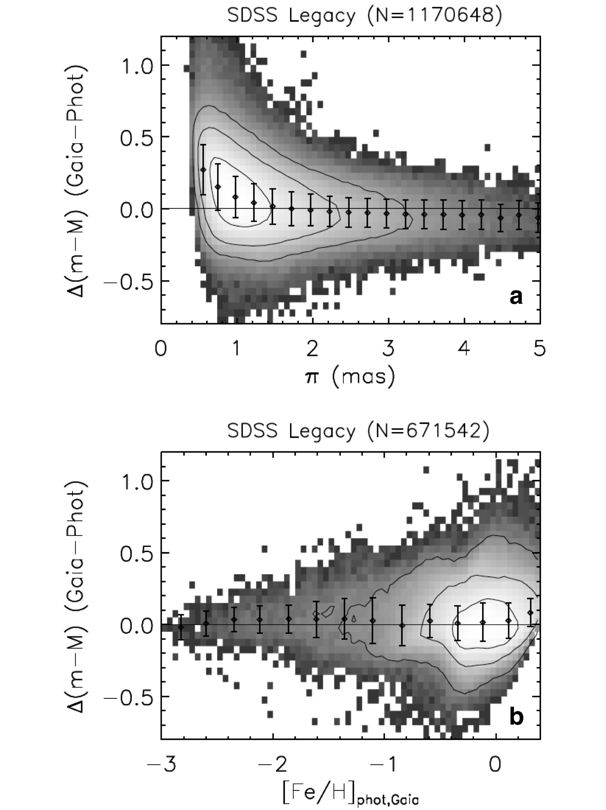}
\caption{Comparison between photometric distances and Gaia parallaxes. Comparisons are shown as a function of parallax in panel (a) and photometric metallicity with Gaia priors in panel (b). In panel (b), only stars with large parallaxes ($\pi > 1$~mas, i.e., more nearby) are included in the comparison. The gray scale shows the logarithmic number density of the stars; the contours indicate 1, 10, and 50 percentiles. The points with error bars indicate moving averages and standard deviations. The sample includes all stars in the SDSS Legacy Survey region with $\sigma_\pi/\pi < 0.20$.}
\label{fig:parallax} \end{figure}

Figure~\ref{fig:parallax} shows a comparison between the photometric and Gaia distances. The latter values have been corrected for the zero-point offset in Gaia \citep{lindegren:18}. Overall, the agreement is satisfactory, and distance estimates from the full photometric solutions are within 2\% of the Gaia parallaxes over a wide range of metallicities for nearby stars ($\pi > 1$~mas).

\subsection{Rotational Velocities}

An accurate measurement of rotational velocities in the Galactocentric cylindrical coordinate system requires a full three-dimensional motion of a star, which is calculated from a heliocentric radial velocity and a proper-motion measurement. However, most of the stars included in this work lack radial-velocity measurements. Therefore, we computed rotational velocities based on distances and proper motions of individual stars along the Prime Meridian in the Galaxy, which is perpendicular to the direction of motion in the Galactic disk (see Figure~\ref{fig:map}).

We adopted $8.34$~kpc for the distance to the Galactic center \citep{reid:14}, the Sun's velocity with respect to the Local Standard at Rest, $(U, V, W) = (11.1, 12.24, 7.25)\ {\rm km\ s}^{-1}$, and the circular velocity of the LSR being $238~{\rm km\ s}^{-1}$ in the Galactocentric rest frame \citep{schonrich:12}. We corrected the rotational velocity derived from proper motions, $v_{\rm ppm}$, for the geometric inclination effect, using $v_{\rm proj} = v_{\rm ppm}$ sec $\zeta$, where $\zeta$ is the angle measured from the Galactic Prime Meridian and $v_{\rm proj}$ is the projected rotational velocity. We note that, within $\pm30$ degrees of the Galactic Prime Meridian, the difference between our inferred rotation velocity and the actual rotation velocity is, at most, about 1.5\% at $\pm10$ degrees from the Meridian, and about 15\% at $\pm30$ degrees from the Meridian.

\subsection{Effects of Contaminants}

Our photometric metallicity estimates assume that all stars are main-sequence dwarfs. However, the separation between dwarfs and giants is non-trivial from conventional broad-band photometry alone. Unrecognized giants appear to be more metal rich, thus distant giants in the survey data can bias metallicity distributions in local volumes. Fortunately, the fraction of distant giants in a given local volume is only on the order of 10 percent \citep{juric:08,an:13}, although it is a strong function of stellar color ranges and/or the sample selection scheme. The effects are most pronounced for a nearby volume at $|Z| < 2$~kpc, due to the large number of distant giants in the halo, but the contamination rate is less than 5\% at larger distances ($|Z| > 3$~kpc), if one assumes a significantly reduced number of stars beyond $\sim30$~kpc from the Galactic center \citep{sesar:10}.

Unresolved binaries are another potential source of contamination in the photometric metallicity mapping. The fraction and the exact form of the mass function for secondary stars are not well-known for field halo stars, which are important factors in determining the degree of bias. If the binary fraction and the mass distribution function are similar to those in the Solar Neighborhood \citep{duquennoy:91}, or in globular clusters \citep{milone:12}, most of binaries in our halo sample cannot be distinguished from single-epoch images in SDSS or SCUSS. They appear more metal poor than single stars by a few tenths of dex \citep{an:13}, and appear closer due to the combined light from two sources. The resulting effect is a blurring of the sample distribution in the distance versus metallicity plane, although the metallicity bias should be negligible for low mass-ratio binaries. Nevertheless, our previous analysis from artificial star tests showed that the overall metallicity bias is dominated by typical photometric errors, not by the fraction of binaries in the sample \citep{an:13}.

\section{Results}\label{sec:results}

\subsection{[Fe/H] vs. $v_\phi$ Distribution}

\begin{figure*}
\centering
\includegraphics[scale=0.80]{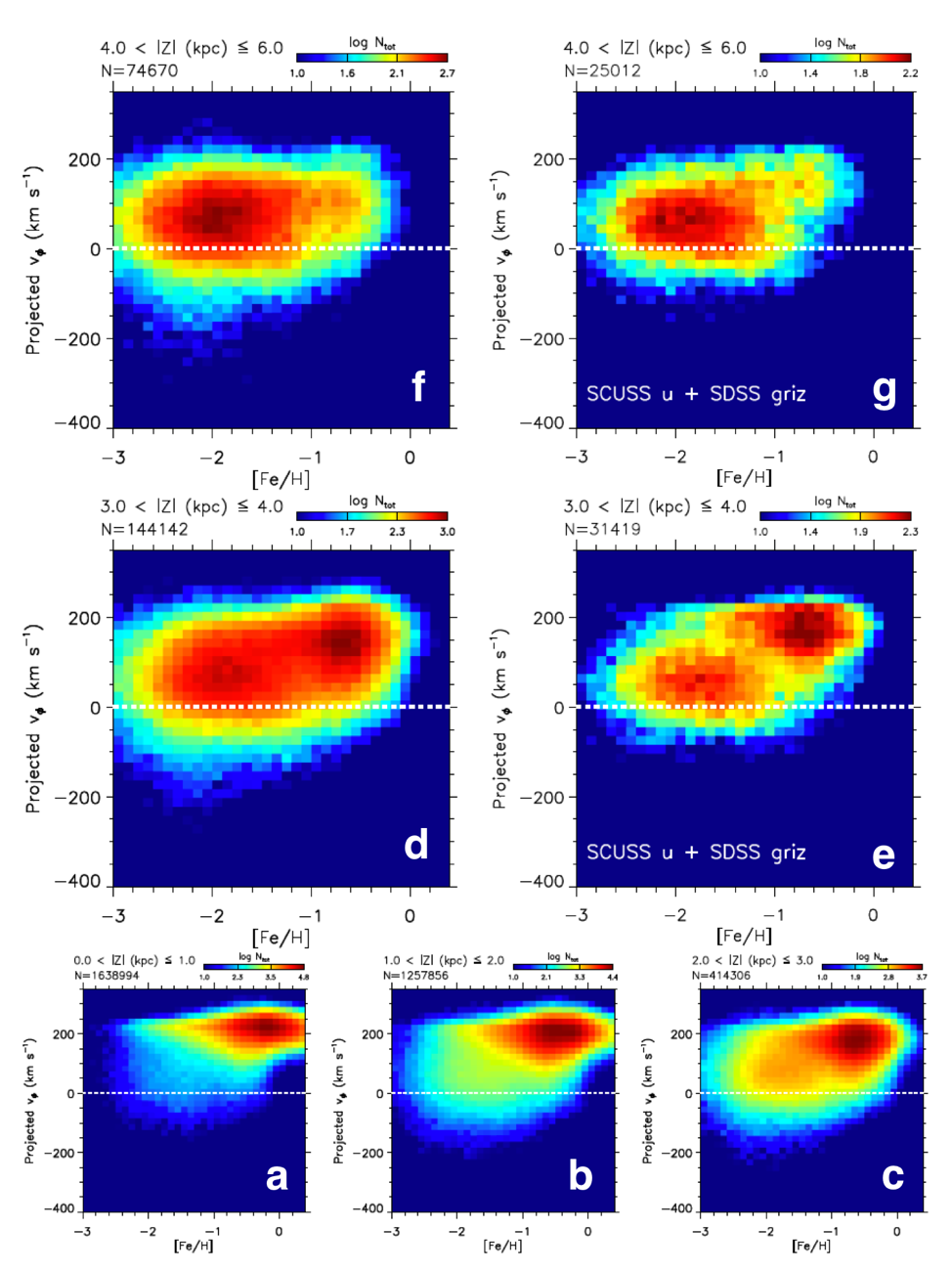}
\caption{Distribution of the sample in a multi-dimensional parameter space. Logarithmic number distributions are shown in a rotational velocity versus metallicity space for different slices of the vertical distance from the Galactic plane. Distances and metallicities are obtained using SDSS photometry, and rotational velocities in the Galactocentric cylindrical coordinate system are computed from Gaia proper motions and photometrically derived distances along the Galactic Prime Meridian. The rotational velocities are corrected for inclination with respect to the Prime Meridian. The horizontal dashed line indicates zero rotation in the Galactocentric rest frame. Near the Galactic plane ($|Z| < 2$~kpc), kinematically hot stars (those falling in the range $-100$~km s$^{-1} < v_\phi < 100$~km\ s$^{-1}$) have a mean metallicity of [Fe/H]$\sim-1.5$, which is consistent with previous studies of metal-poor halo stars. However, the mean value shifts toward lower metallicity at large vertical distance. Panels (e) and (g) show the same distribution of stars as in panels (d) and (f), respectively, but based on SCUSS $u$-band photometry rather than SDSS $u$-band photometry. The improved metallicity estimates and depth of the SCUSS data result in a clear separation between the disk and the halo stellar components, even with a factor of $3$--$4$ fewer stars.}
\label{fig:sample} \end{figure*}

Figure~\ref{fig:sample} shows the distribution of our sample in [Fe/H] versus $v_\phi$, as a function of distance from the Galactic plane ($Z$). Near the Galactic plane, disk stars show a `$>$'-shaped distribution, which reflects negative and positive metallicity-velocity correlations found among thin-disk and TD stars, respectively \citep{lee:11,belokurov:19}. At $|Z| > 3$~kpc, a clear separation between the traditional disk and halo populations is seen, all of which verify our photometric approach as a distance and metallicity estimator. We note that, in all distance bins, most stars in our sample are found in prograde rotation, in the same sense as the Sun's motion around the Galactic center.

\begin{figure*}
\centering
\includegraphics[scale=0.80]{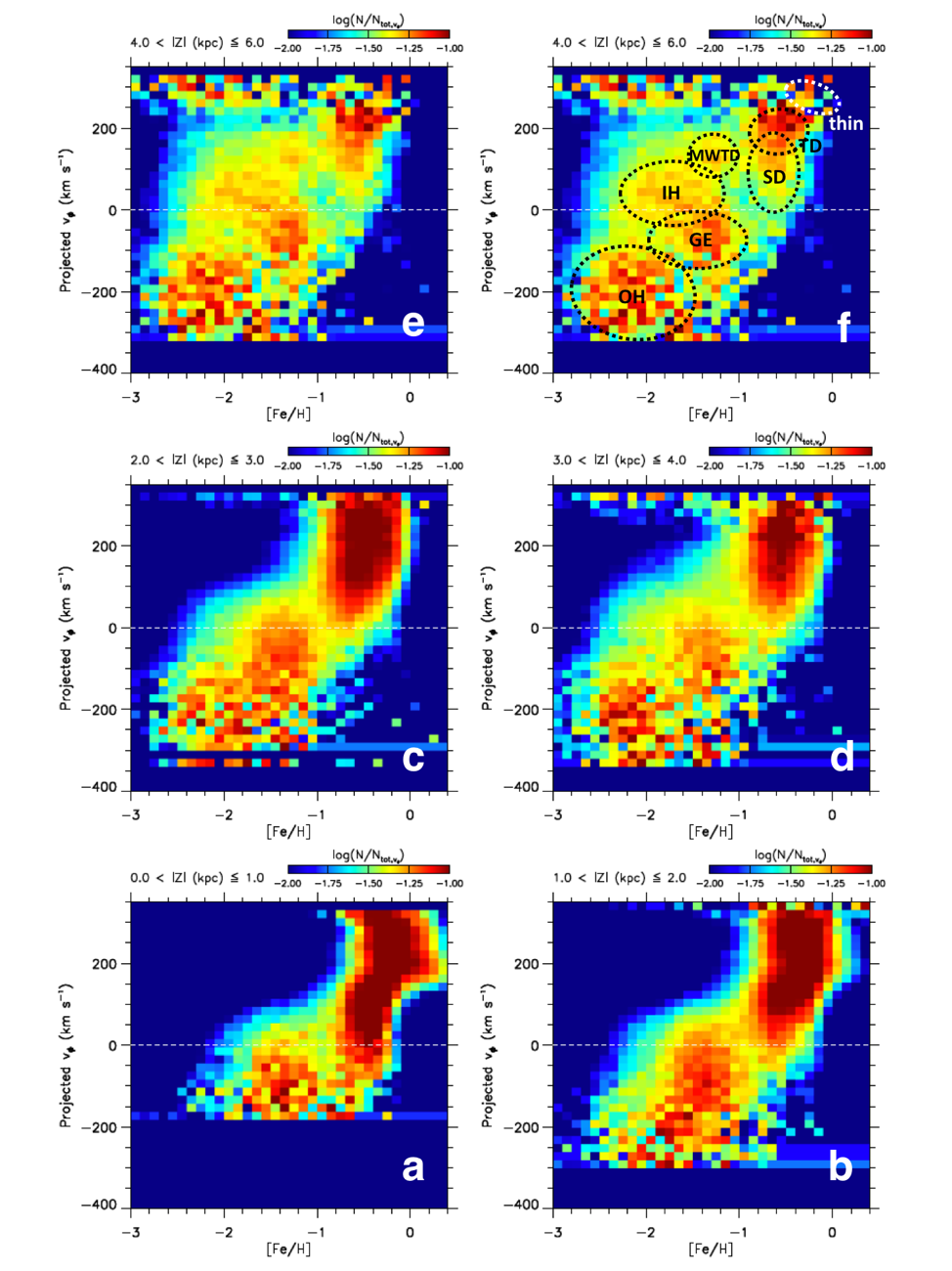}
\caption{Normalized metallicity distribution as a function of rotational velocity. The same set of stars are used as in Figure~\ref{fig:sample}, but the number of stars in each pixel, weighted by photometric metallicity errors, is normalized with respect to the total number of stars in each $v_\phi$ bin, and shown in logarithmic number ratios. Only pixels with more than 50 stars in each $v_\phi$ bin are displayed. The horizontal dashed line indicates zero rotation in the Galactocentric rest frame. The dominant halo component, centered at ${\rm [Fe/H]} \approx -1.4$ and $v_\phi \approx -50$ km s$^{-1}$ is observed throughout the volume, but a more metal-deficient component with an extended distribution in the parameter space (centered at ${\rm [Fe/H]} \approx -2.2$ and $v_\phi \approx -250$~km\ s$^{-1}$ at $|Z| > 3$~kpc) becomes increasingly visible at large distances. At $|Z| > 4$~kpc (panel e), where the contribution from disk stars is minimal, additional clumps begin to reveal themselves: the broad component centered at ${\rm [Fe/H]} \approx -1.6$ and $v_\phi \approx +50$ km s$^{-1}$, and a small clump at ${\rm [Fe/H]} \approx -1.3$ and $v_\phi \approx +150$ km s$^{-1}$. Panel (f) marks the positions of various stellar components in the Galaxy: Gaia-Enceladus (GE), the inner-halo (IH), the outer-halo (OH), the canonical thick disk (TD), the metal-weak thick disk (MWTD), and the Splashed Disk (SD).  Note that the MWTD is clearly an independent structure from the TD, as suggested by several recent spectroscopic analyses \citep{beers:14,carollo:19}. An approximate location of the thin disk, which is not visible in this distance bin, is shown by a white dotted ellipse.}
\label{fig:blueprint} \end{figure*}

However, it is noteworthy that the number-density distributions in Figure~\ref{fig:sample} are dominated by the major structures (disk and halo), and the presence of small clumps of stars in the [Fe/H]-$v_\phi$ space are swamped by the large dynamic range of the number density shown in this figure. For this reason, in Figure~\ref{fig:blueprint} we normalized the number-density distribution along each $v_\phi$ ($N/N_{\rm tot,v_\phi}$), more effectively displaying the metallicity distribution of the stars as a function of $v_\phi$. Here, the number count of stars in each pixel was weighted by photometric metallicity errors ($1/\sigma_{\rm [Fe/H]}^2$).

The normalized chemo-dynamical distribution of stars in Figure~\ref{fig:blueprint} appears dramatically different from those in Figure~\ref{fig:sample} -- detailed substructures emerge, especially in the low number-density regions with negative $v_\phi$. The most striking substructure is a clump of stars centered at ${\rm [Fe/H]} \approx -1.4$ and $v_\phi \approx -50$~km s$^{-1}$, which we assign to GE. The GE component was originally recognized as a major retrograde structure, but was considered to include stars in more extreme retrograde orbits. However, our metallicity map indicates that GE is clearly separated from an extended distribution of stars with large retrograde motions ($\langle v_\phi \rangle \approx -200$ km s$^{-1}$) at very low metallicities ($\langle {\rm [Fe/H]} \rangle \approx -2.2$), which we assign to the OH component \citep{carollo:07}. Near GE, there is a low signal-to-noise clump at ${\rm [Fe/H]} \approx -1.2$ and $v_\phi \approx -150$ km s$^{-1}$. The large retrograde motions of the stars in this clump indicate its potential connection to several previously suggested retrograde structures, possibly related to the Sequoia Event \citep{koppelman:18,myeong:19}, but it is not clear whether they are more closely related to the OH component.

At $|Z| > 4$~kpc, where the contribution from numerous disk stars is minimal, a number of features that can be matched to those identified in previous studies -- the IH, MWTD, and the SD -- are readily visible on the map. Among them, a clump in prograde motion centered at ${\rm [Fe/H]} \approx -1.6$ and $v_\phi \approx 50$~km s$^{-1}$ shows a more extended metallicity distribution than GE; we assigned this to a superposition of the IH and OH components (see below).

\begin{figure}
\centering
\includegraphics[scale=0.86]{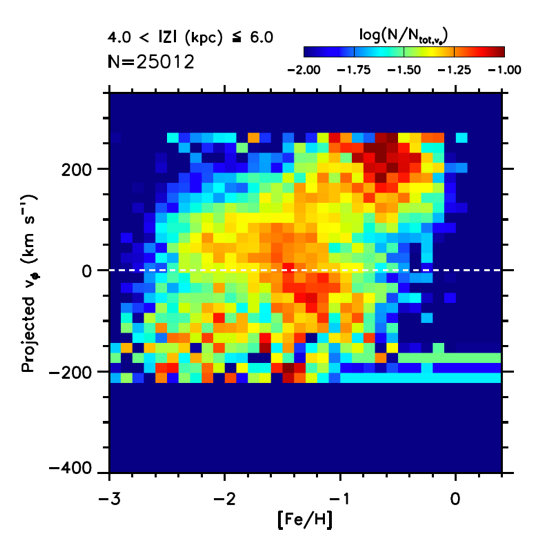}
\caption{Normalized metallicity distribution from SCUSS.  This is same as in panels (b) and (c) in Figure~\ref{fig:blueprint}, but is based on the SCUSS $u$-band photometry. Only pixels with more than 50 stars in each $v_\phi$ bin are displayed. The horizontal dashed line indicates zero rotation in the Galactocentric rest frame. The total number of stars included in the above plot is about a factor of three smaller than the case obtained using the SDSS $u$-band ($N = 74,670$), leading to a weaker signal.}
\label{fig:scuss_blueprint} \end{figure}

The same result is obtained from the combined SCUSS/SDSS catalog. Although the signal is lower due to the smaller number of stars in the sample, SCUSS provides more precise $u$-band measurements than SDSS, and therefore can be used to obtain more precise metallicity estimates from photometry. Figure~\ref{fig:scuss_blueprint} is the same version of a normalized metallicity distribution of stars at $4 < |Z| ({\rm kpc}) < 6$ as in Figure~\ref{fig:blueprint}. Although SCUSS has higher precision photometry, it covers a significantly smaller region along the Galactic Prime Meridian. As a result, the signal-to-noise ratio in the SCUSS map is lower. Nevertheless, the major Galactic components -- GE, the IH component (in combination with the OH component), the canonical TD, and the MWTD -- can be identified from Figure~\ref{fig:scuss_mdf}, although the outer-halo component is significantly weaker.

\begin{figure}
\centering
\includegraphics[scale=0.86]{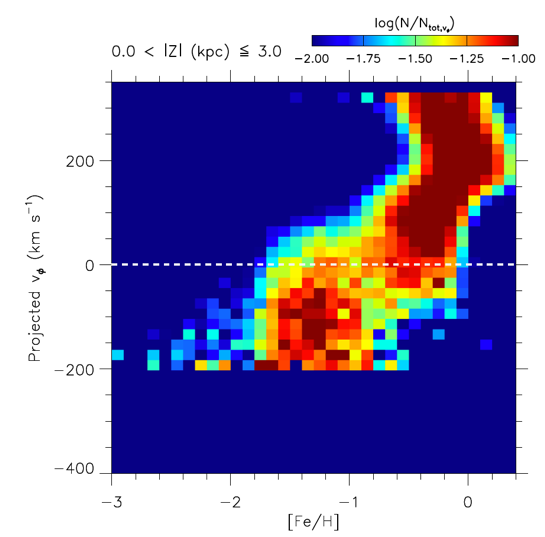}
\caption{Normalized metallicity distribution with Gaia priors. The same as in Figure~\ref{fig:blueprint}, but based on metallicity estimates using Gaia priors on distance. Only pixels with more than 50 stars in each $v_\phi$ bin are displayed. The horizontal dashed line indicates zero rotation in the Galactocentric rest frame.  The sample ($N=875,570$) includes stars with good trigonometric parallaxes ($\sigma_\pi/\pi < 0.2$).}
\label{fig:gaia_blueprint} \end{figure}

For nearby stars, we combined the adopted photometry with an individual stas's parallax from the Gaia mission ($\sigma_\pi/\pi < 0.2$) for improved precision in the derived metallicity. This approach provides at least a factor of two improved precision in the metallicity estimation (see Figure~\ref{fig:param}). However, good parallaxes from Gaia are limited to nearby stars, and a cross-match with our photometric samples results in a maximum vertical distance of $|Z|\approx 3$~kpc for its application. Figure~\ref{fig:gaia_blueprint} shows a normalized metallicity distribution of stars, based on Gaia priors on individual stellar distances. Because the volume is limited by numerous disk stars, most of the halo substructures, except GE, are not visible in this plot.

\subsection{[Fe/H] vs. $|Z|$ Distribution}

\begin{figure*}
\centering
\includegraphics[scale=1.0]{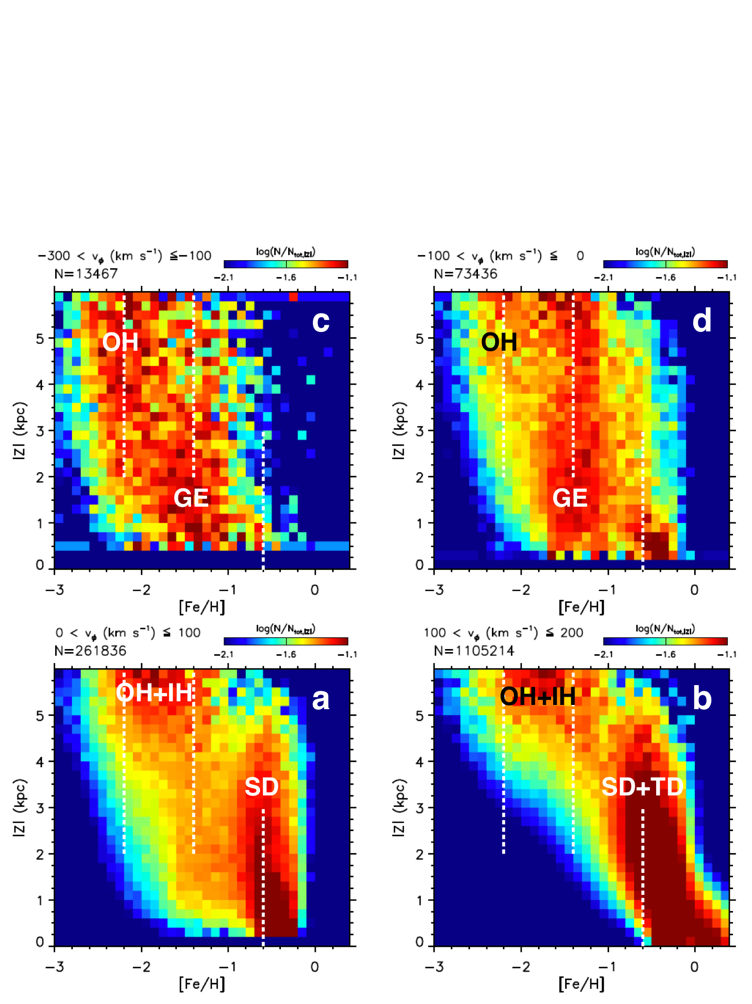}
\caption{Normalized metallicity distribution as a function of vertical distance from the Galactic plane. The number of stars in each pixel, weighted by photometric metallicity errors, is normalized with respect to the total number of stars in each distance bin, and displayed in a logarithmic number ratio in four different velocity bins. Only pixels with more than 30 stars in each $v_\phi$ bin are displayed. The range of the rotational velocity is chosen to isolate the individual stellar components from $4 < |Z|$~(kpc)$ < 6$ in Figure~\ref{fig:blueprint}. The vertical dotted lines (at [Fe/H] = $-0.6$, $-1.4$, and $-2.2$) are overlaid to guide the eye.}
\label{fig:vertical} \end{figure*}

The vertical structure of the metallicity distribution is shown in Figure~\ref{fig:vertical}, in four different slices of $v_\phi$, in order to better isolate each of the major components. In the most retrograde bin, the halo is dominated by metal-poor OH stars at larger distances above the Galactic plane. More metal-rich GE debris occupies the volume near the Sun at $|Z| < 3$~kpc, among those with retrograde motions. For stars with mild retrograde motions, there appears a bottleneck in the metallicity distribution at $|Z|\approx4$~kpc, probably marking a transition zone from the flattened GE debris to more spatially extended structures -- a combination of the IH and OH components. At $|Z| > 4$~kpc, a smooth transition occurs from low-metallicity ($\langle {\rm [Fe/H]} \rangle \approx -2.2$) stars at $v_\phi < -100$~km~s$^{-1}$ to a more extended metallicity distribution ($-2.2 < {\rm [Fe/H]} < -1.4$) with prograde motion. The majority of these stars constitute either the IH or OH components.

\subsection{Systematic Change in the Metallicity Distribution}

\begin{figure*}
\centering
\includegraphics[scale=1]{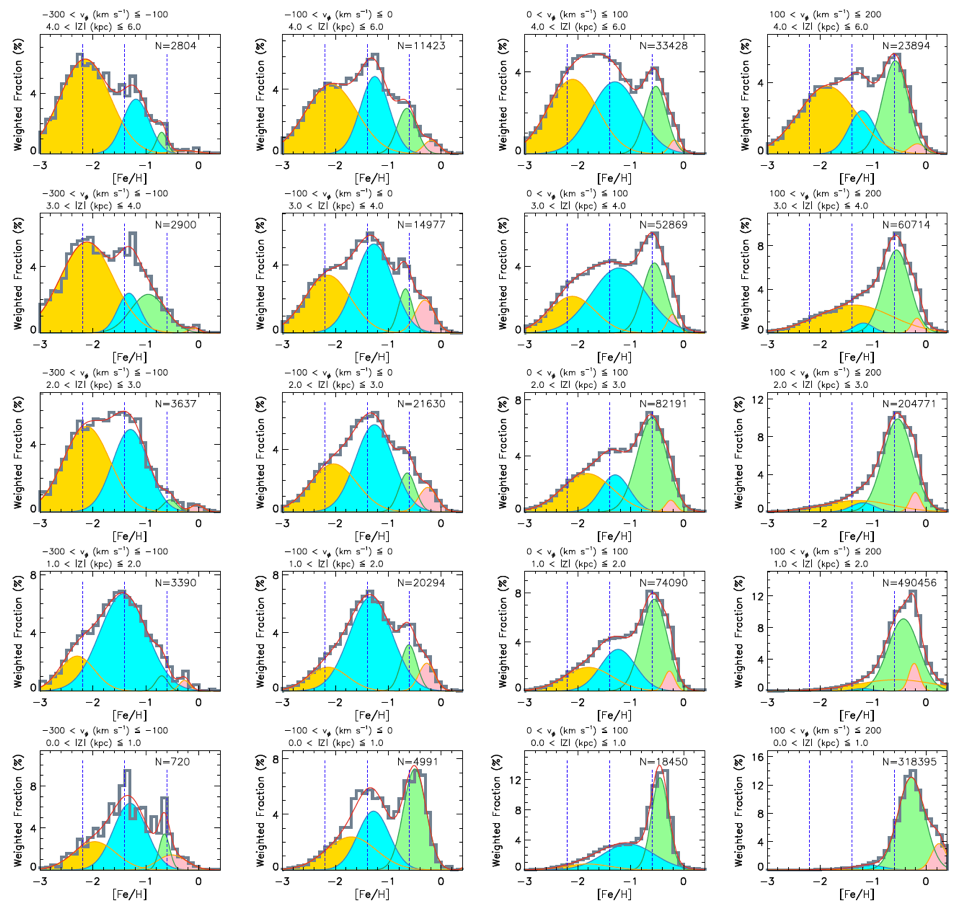}
\caption{Substructures in the metallicity distribution. Panels are arranged as a function of rotational velocity (from the left to the right panels) and vertical distance from the Galactic plane (from the bottom to the top panels) along the Galactic Prime Meridian. The number of stars in each metallicity bin is weighted by their metallicity errors. The observed distribution is well-fit by four different normal distributions, each representing one or more of the major components of the Galaxy: the OH (yellow shaded), the IH and/or GE (blue shaded), the TD, MWTD, and/or the SD (green shaded), and the thin disk and/or TD (red shaded). The gray histogram is the sum of all of the data; the red solid line represents the sum of the fit components.  The most metal-rich component (red shaded) may have additional contributions from distant giants in the halo. Decomposition of the observed distribution is only intended to show approximate mean locations and extensions of the various stellar populations. The vertical dotted lines (at [Fe/H] = $-0.6$, $-1.4$, and $-2.2$) are overlaid to guide the eye.}
\label{fig:mdf} \end{figure*}

In Figures~\ref{fig:sample} and \ref{fig:vertical}, halo stars are under-represented near the Galactic plane, since they are simply swamped by the significantly more numerous disk stars. A more quantitative evaluation of each component's contribution is given in Figure~\ref{fig:mdf}, which displays the metallicity distribution for stars in slices of $v_\phi$ and $|Z|$. At $|Z| >$~3 kpc, stars with retrograde rotation exhibit a metallicity distribution with characteristic double peaks at ${\rm [Fe/H]} \approx -2.2$ and $\approx-1.4$, reflecting approximately equal contributions from the OH component and a mixture of the GE and IH stars. The duality of the halo was originally inferred from a gradual shift in the metallicity distribution as a function of rotational velocity and vertical distance in a local kinematic analysis \citep{carollo:07}, and confirmed by in-situ evidence for a radial metallicity gradient based on models of Hess diagrams derived from SEGUE photometry \citep{dejong:10}, and by numerous  analyses since. Although double peaks in the metallicity distribution of metal-poor halo stars were hinted at in recent spectroscopic analyses \citep[e.g.,][]{fernandez:17,conroy:19}, the discrete peaks and their systematic variations in Figure~\ref{fig:mdf} are a direct proof of the presence of multiple stellar populations in the halo. Photometric errors and unresolved binaries are unlikely to bias metallicity estimates, since stars in the same volume, but with prograde rotation, do not exhibit multiple peaks in their metallicity distributions.

In general, a normal distribution is not adequate to describe the observed metallicity distribution in [Fe/H], because even the simplest stellar population in a closed-box system is expected to show a low-metallicity tail \citep[see][]{ryan:91}. The tail may even appear more pronounced due to larger photometric errors at low metallicity \citep{an:13}. Nevertheless, it is useful to decompose the observed distribution using a set of normal distributions to evaluate the varying (approximate) contributions of the individual components \citep[e.g.,][]{ivezic:08}. In Figure~\ref{fig:mdf}, shaded curves are the best-fit set of normal distributions in each distance and velocity bin. We employed four different components in the fit, which are the minimum number required to capture the observed distributions in all panels. We kept the same color scheme for each component to track its slowly varying contribution and centroid, but each fitted component does not necessarily match each of the above-mentioned stellar populations.

At $100\ {\rm km\ s}^{-1} < v_\phi < 200\ {\rm km\ s}^{-1}$, the MWTD emerges, with significantly weaker fractions than the halo components, which are fit by two normal distributions. Similarly, the SD population is not readily distinguishable from the classical TD population because of their similar metallicities. If one follows its fitted distribution, however, it can be seen that it extends far above the Galactic plane and to retrograde motions, properties that are consistent with those obtained from previous spectroscopic work \citep{belokurov:19}. GE and the IH component, although they are shown as separate entities in Figures~\ref{fig:blueprint} and \ref{fig:vertical}, are also fit using the same colored normal distribution due to their mild metallicity difference.

According to the above fitting exercise, about two thirds of metal-poor stars at $1 < |Z| < 2$~kpc belong to GE, if one restricts the sample to $v_\phi < 100\ {\rm km\ s}^{-1}$, but the fraction decreases to about $40\%$ at $2 < |Z| < 3$~kpc. At large vertical distances ($4 < |Z| < 6$~kpc), the ratio of the IH and OH components is near unity, in agreement with our previous work based on a more-restricted sample \citep{an:13}.

\begin{figure*}
\centering
\includegraphics[scale=0.8]{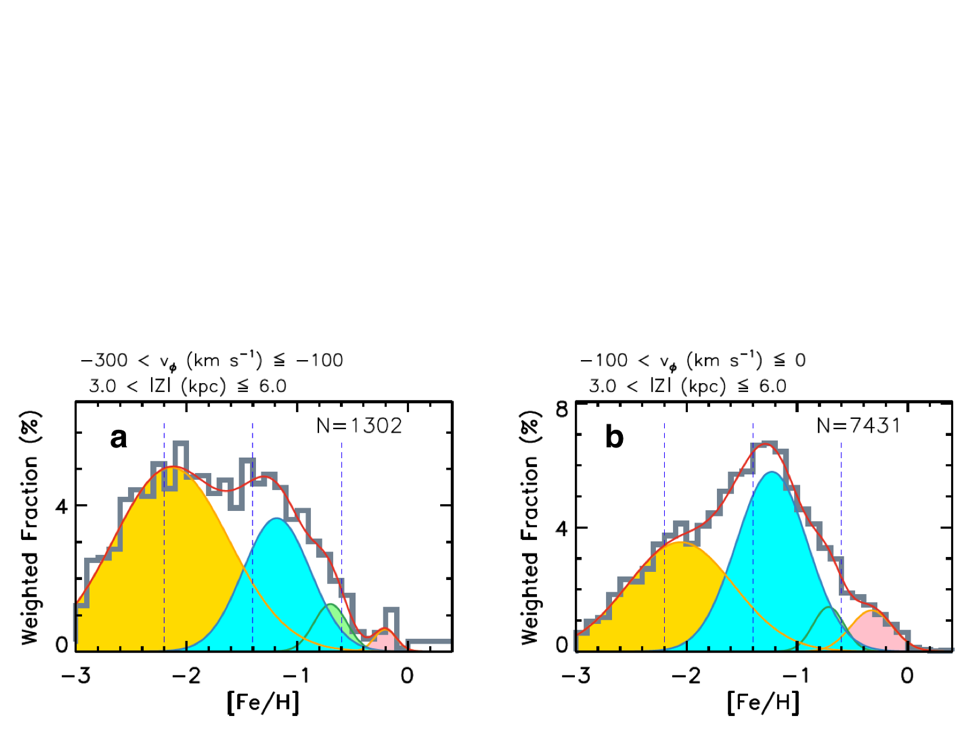}
\caption{Metallicity distributions from SCUSS $u$-band data. This is same as in Figure~\ref{fig:mdf}, but the distributions are derived from distance and metallicity estimates using the SCUSS $u$-band photometry. The observed distribution is well-fit by four different normal distributions, each representing one or more of the major components of the Galaxy: the OH (yellow shaded), the IH and/or GE (blue shaded), the TD, MWTD, and/or the SD (green shaded), and the thin disk and/or TD (red shaded). The gray histogram is the sum of all of the data; the red solid line represents the sum of the fit components. The vertical dashed lines (at [Fe/H] = $-0.6$, $-1.4$, and $-2.2$) are overlaid to guide the eye.}
\label{fig:scuss_mdf} \end{figure*}

\begin{figure*}
\centering
\includegraphics[scale=0.8]{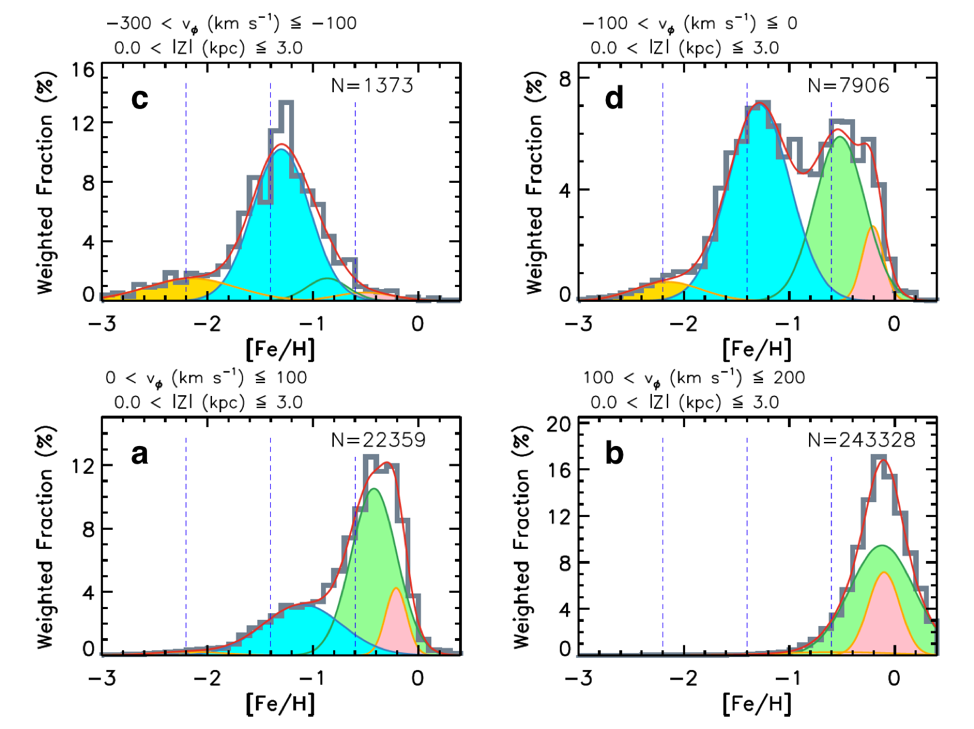}
\caption{Metallicity distributions with Gaia priors. This is same as in Figure~\ref{fig:mdf}, but for a subset of the metallicity distributions at $|Z| < 3$~kpc with photometric metallicity estimates with Gaia priors on distance.  The observed distribution is well-fit by four different normal distributions, each representing one or more of the major components of the Galaxy: the OH (yellow shaded), the IH and/or GE (blue shaded), the TD, MWTD, and/or the SD (green shaded), and the thin disk and/or TD (red shaded). The gray histogram is the sum of all of the data; the red solid line represents the sum of the fit components. The vertical dashed lines (at [Fe/H] = $-0.6$, $-1.4$, and $-2.2$) are overlaid to guide the eye.  Only stars with good parallaxes from Gaia ($\sigma_\pi/\pi < 0.2$) are shown.}
\label{fig:gaia_mdf} \end{figure*}

Metallicity distributions from photometric estimates with SCUSS photometry, or those with Gaia priors on distance, also support the discrete nature of the individual stellar populations. As shown in Figure~\ref{fig:scuss_mdf}, the metallicity distributions of stars in retrograde motions from SCUSS clearly support two metal-poor components. Stars with good Gaia parallaxes are limited to a local volume ($|Z| < 3$~kpc). Nevertheless, the observed metallicity distributions with Gaia parallaxes in Figure~\ref{fig:gaia_mdf} show a characteristic kink at ${\rm [Fe/H]} \approx -2$, indicating a trace contribution of the OH component near the Galactic plane.

\section{Discussion}\label{sec:discussion}

Our chemo-rotational mapping helps to visually identify and delineate the extension of each of the stellar components to an unprecedented level of completeness. Earlier studies on the duality of the smooth halo \citep{carollo:07,carollo:10,beers:12} may have included a large fraction of the GE (unknown at that time) stars in their IH sample, because of their similar metallicities and rotational velocities, leading to a systematically higher mean metallicity and a smaller net rotational velocity. If the IH and GE are separate entities, as our map supports, the actual spatial distribution of IH stars may be less centrally concentrated, and less oblate than previously envisaged. Spectroscopic studies also indicated that metal-poor halo stars on low-eccentricity orbits exhibit high $\alpha$-element abundances with respect to iron \citep{nissen:10,hayes:18,mackereth:19}, while stars in GE on strongly radial orbits show lower $\alpha$-element abundances \citep{helmi:18}. This suggests a more active star-forming environment for stars associated with the IH component, possibly having formed in massive gas clumps in the central region of the proto-Galaxy, and subsequently displaced to their current higher-energy orbits by scattering processes \citep{zolotov:09,garrison:18}. The chemical enrichment of the IH component (which should now be referred to specifically as an in-situ halo, in contrast to the accreted components) may have influenced the formation of the metal-weak and/or the classical thick disks.

On the other hand, the low metallicity of the OH component, along with its low $\alpha$-element abundances \citep{matsuno:19} indicates that it may have originated from the accretion of low-mass dwarf galaxies,\footnote{These ideas were speculated upon as early as two decades ago, based on the analyses by \citet{sommerlarsen:97} and \citet{chiba:00}.} which have been merged into the proto-Galaxy in accordance with theoretical predictions of galaxy formation \citep{bullock:05,tissera:13}. Ongoing and future photometric surveys such as Pan-STARRS \citep{chambers:16}, SkyMapper \citep{wolf:18}, J-PLUS \citep{cenarro:19}, S-PLUS \citep{mendes:19}, and LSST \citep{ivezic:19}, as well as larger and more accurate astrometric catalogs, are promising assets to explore a larger and finer parameter space of stellar populations in the Galactic halo.

\acknowledgements

We thank an anonymous referee for comments that clarified our presentation. We also thank Young Sun Lee for his useful comments on the SSPP. D.A. acknowledges support provided by Basic Science Research Program through the National Research Foundation of Korea (NRF) funded by the Ministry of Education (NRF-2018R1D1A1A02085433) and by the Korean NRF to the Center for Galaxy Evolution Research (No.\ 2017R1A5A1070354). T.C.B. acknowledges partial support from grant PHY 14-30152 (Physics Frontier Center/JINA-CEE), awarded by the U.S. National Science Foundation. Both authors acknowledge the use of SDSS data (https://www.sdss.org/).

{}

\end{document}